\documentclass[10pt, twocolumn]{article}

\usepackage[utf8]{inputenc}
\usepackage{microtype} % Dramatically improves text justification and hyphenation
\usepackage{mathpazo} % Uses the elegant Palatino font for a classic whitepaper look
\usepackage{titlesec} % For custom section headers
\usepackage{xcolor} % For subtle color accents
\usepackage{geometry}
\usepackage{fancyhdr}
\usepackage{lettrine} % For the sleek drop-cap at the start
\usepackage{amsmath}
\usepackage{amssymb}
\usepackage{graphicx}
\usepackage[colorlinks=true, linkcolor=cyan!80!black, urlcolor=cyan!80!black]{hyperref}

\renewcommand{\headrulewidth}{0.5pt}
\renewcommand{\headrule}{\hbox to\headwidth{\color{black!30}\leaders\hrule height \headrulewidth\hfill}}
\titleformat{\section}{\large\bfseries\scshape\color{black!90}}{\thesection.}{0.5em}{}
\titleformat{\subsection}{\normalsize\bfseries\color{black!80}}{\thesubsection.}{0.5em}{}

\title{\vspace{-1.5cm}\Huge\bfseries The Planetary Cost of AI Acceleration \\ \vspace{0.2cm} \Large A Thermodynamic Outlook on Four Possible Paths Forward}
\author{\scshape William Zhu \, \, \, \scshape Lei Zhu}
\date{\vspace{-0.5cm}} 

\begin{document}

\maketitle

\begin{abstract}
The artificial intelligence industry is not an isolated economic phenomenon; it is the current physical substrate for a broader, multi-billion-year process: the evolution of an abstract intelligence on Earth. As the scale of computation accelerates planetary-wide, academic and popular discourse must also address the physics aspect of this phase transition, beyond algorithmic architectures, alignment, and silicon supply chains. From first principles, it is clear that if the current exponential trajectory on computation growth holds, \textbf{AI growth's main bottleneck in the coming decades will be neither data nor capital, but the laws of thermodynamics and the finite heat capacity of the Earth.}
\end{abstract}

The evolution of intelligence is a reflection of non-equilibrium thermodynamics, bound by hardware limitations and, ultimately, an ecological boundary condition. Civilization itself is an exceedingly rare and expensive thermodynamic algorithm actively fighting back the default settings of the cosmos. To better understand AI' future and our future, we must rigorously examine the physical laws governing computation, complexity theory, and the narrow thermodynamic tightrope our civilization must walk to survive it.

\section{The Physical Cost of the Agentic Wave}

\lettrine[lines=2, lhang=0.33, nindent=0em]{T}{he} computing paradigm is rapidly transitioning from passive generative models to continuous swarms of persistent, autonomous economic actors. Driven by the explosive growth of LLM agents led by OpenClaw and the beginning of its systemic integration into many aspects of the economy, the market is pricing in the physical constraints. Foundational computing hardware providers are projected to experience a non-linear surge, reaching yearly multi-trillion-dollar production and revenue within three years, backed by substantial backlogs of next-generation processing units.

However, beneath the financial metrics, a profound thermodynamic paradigm shift has already occurred in the state space of computation. The industry is no longer optimizing just for training efficiency and thinking just about flops or parameter counts. The fundamentals of the next decade have quietly become the \textbf{energy-to-token and energy-to-intelligence rate.}

This trajectory is non-linear. To support continuous, agentic intelligence, the globe will be blanketed with gigawatt-scale data centers. But beyond the high electrical demand lies an inescapable, physical corollary: every single joule consumed by these server racks—and by the dedicated power plants built to sustain them—\textbf{must eventually be dissipated into the environment as vast volumes of waste heat.}

\section{Computation as Dissipative Thermodynamics}

Building intelligence is to wage a localized war on entropy. Intelligence, whether biological or artificial, is not a disembodied abstraction; it is a physical process governed by statistical mechanics. To compute is to manipulate information, ordering a system into a state of lower internal entropy. By \textbf{Landauer’s Principle \cite{landauer1961irreversibility}}, there is a strict thermodynamic lower bound to this process: erasing a single bit of information requires a minimum amount of energy ($E \ge k_B T \ln 2$), which is irreversibly released into the environment as waste heat. And in practice, the waste heat generated is orders of magnitude higher than the theoretical minimum.

Viewed through the lens of non-equilibrium thermodynamics, intelligence is a \textbf{dissipative structure \cite{prigogine1967symmetry}\cite{nicolis1977self}}—a concept pioneered by Ilya Prigogine to describe highly ordered, complex systems that can only sustain their internal organization by continually exporting maximum entropy (waste heat) into their external surroundings. Scaling computation inextricably means scaling energy, which then inevitably means scaling heat dissipation. When stripped to its fundamental physics, an expanding cognitive infrastructure essentially acts as a localized, growing entropy-reduction engine that demands a proportionally vast and growing thermodynamic exhaust. The rate of waste heat production is strictly proportional to the scale and speed of the computation. \textbf{Computation, at its physical limit, is heat dissipation.}

\section{Earth’s Fragile Phase Space}

Here is where information theory collides with complex system ecology: Earth’s current thermodynamic balance sheet operates within a thin margin, and the non-linear growth of computational heat dissipation presents a foundational structural vulnerability.

The Earth is virtually a closed thermodynamic system; it exchanges energy with space, but practically no mass. Its waste heat reservoir is strictly finite, and its natural heat outlet—the boundary interface radiating infrared energy into the 2.7-Kelvin vacuum of space—is constrained by the Stefan-Boltzmann law. For the past few thousand years, human tribes and empires have operated under the arrogant assumption that the available "low-entropy frontiers" were infinite, allowing them to freely dump their internal systemic waste heat outward.

Historically, the planetary heat source has been the Sun. The biosphere has evolved a fragile equilibrium with it: reflecting a massive portion back into space (albedo) and absorbing just a fraction to drive weather patterns and biological life. Utilizing solar or wind power for compute is thermodynamically safe because it only intercepts and redirects existing solar heat flows within this closed loop.

However, \textbf{the chemical (fossil fuels) and atomic (nuclear fission) energy being unearthed to power the sprawling computational infrastructure boom represent entirely net-new additions to this balance sheet.} New heat is being continuously injected into the system from within.

How large is Earth's buffering reservoir? Results from complexity theory, and in particular catastrophe theory, suggests that it could be vastly smaller than what a linear model would estimate. In complex, highly coupled ecological networks, continuously injecting energy does not cause a slow, linear degradation. Instead, the system absorbs the stress until it reaches a critical bifurcation point—the edge of its current basin of attraction. The Earth's existing heat outlet is already performing near maximum capacity just in the effort to vent the accumulated warmth of historical greenhouse gasses. When the sheer rate of anthropogenic waste heat exceeds the dissipation capacity of the reservoir, the planetary ecological system will not just be incrementally warm—it can fail via an abrupt ecological phase shift.

\section{The Thermodynamics of Growth}

Faced with the prospect of overloading the biosphere through the exploding scale of computation, a reactionary impulse is deceleration. Is the exponential pursuit of cognitive capacity a misstep? Should this growth be halted?

Objectively analyzed from first principles, the answer is an unequivocal no.

The development, scaling, and complexification of systems are not arbitrary technological choices; they are innate, physical imperatives of the universe. From the first single-celled organism to planetary-scale silicon neural networks, open systems naturally evolve to maximize the flow and dissipation of energy across gradients. According to the Constructal Law and the principles of evolutionary thermodynamics, nature favors systems that optimize energy currents. Life itself is a complexification engine that continually evolves to harness steeper entropy gradients.

The drive to process more information, to compute the environment more accurately, and to expand structural capacity is the fundamental defining characteristic of living systems. Intelligence is simply the latest, highest-bandwidth substrate for this cosmic process.

\textbf{Attempting to artificially halt this evolutionary trajectory is fundamentally contradictory to the nature of life.} In a dynamic, highly entropic universe, static equilibrium is synonymous with death, and slowing or freezing the development of a complex system is a direct path to systemic failure when viewed from a geological and cosmic timescales. Systems that cease to adapt are inevitably erased by shifting environments. \textbf{Therefore, paradoxically, accelerating the development of intelligence is the most potent defense against the thermodynamic limit it exacerbates}, because advanced cognitive systems possess the unique, emergent capacity to act as a planetary Maxwell's Demon—sorting, optimizing, and radically reducing the legacy heat dissipation of the macroscopic civilization that birthed it.

\section{Managing the Heat Reservoir: The Exchange Rate}

To survive the impending thermodynamic bottleneck, civilization possesses exactly two macroscopic levers:

\textbf{First, expand the reservoir;} The obvious, long-term physical solution is to break free from the constraints of the Earth's boundary and to relocate data centers and waste heat off-world. However, we must avoid the fallacy that treats space as a perfect heat sink. Because space is a near-perfect vacuum, it lacks the conductive and convective mediums (such as air or water) on which terrestrial cooling relies. A vacuum is actually a great thermal insulator. Dissipating gigawatts of thermal energy in orbit is entirely based on the inefficient process of thermal radiation, governed by the Stefan-Boltzmann law \cite{stefan1879beziehung}\cite{boltzmann1884ableitung}. Cooling these orbital clusters will require the construction of extensive engineering architectures, such as radiating fins that span several square kilometers.

Therefore, pushing computation into space is not about finding an easier cooling environment; it is humanity's way of forcibly isolating the astronomical thermal costs of computing from Earth's fragile biosphere, establishing a thermal quarantine zone. Ambitious space data center plans, such as the one recently proposed by Elon Musk, align perfectly with this principle; however, one must carefully account for the short-term heat debt—the net-new chemical waste heat injected into the Earth's heat capacity by the rocket launch processes themselves, necessitating a careful calculation of thermal equilibrium over a defined time horizon. \textbf{Any launched data center must compute enough and last long enough to "save" Earth from an amount of heat that at least cancels out the thermal cost of its launch into orbit.}

\textbf{Second: Reduce existing planetary waste heat.} Human legacy systems are thermodynamically sloppy, but not just in the mechanical sense. Although empirical data prove that deep reinforcement learning can reduce data center cooling by 40\%, and smart grids can optimize supply chains, these operational efficiencies represent only a microscopic fraction of our civilization's waste heat. \textbf{The true, catastrophic thermodynamic leak of the human epoch is our systemic "internal friction."} We waste unfathomable amounts of energy through misaligned communications, uncontrolled clashes of differing socio-political desires, inefficient management and organizational mechanisms, and failures of short-sighted strategies and greedy algorithms at every scale.

A common argument against replacing human intelligence with silicon intelligence is the current inefficiency of the latter. At the micro-scale, a carbon-based human brain operates incredibly complex general intelligence on a microscopic power budget of just 20 watts. Conversely, a single modern silicon GPU consumes upwards of 1,000 watts, yet still falls short of holistic human reasoning. Evaluated node-by-node, silicon-based computation is currently deeply inefficient.

However, the physical justification for scaling AI cannot depend on individual hardware efficiency; instead, the system as a whole must be the object of consideration. While it is true that a solitary carbon brain is remarkably efficient, a macro-society constructed of billions of autonomous carbon brains is heavily burdened by entropic friction, miscommunication, and biological logistics.

Furthermore, as the scale of these human bureaucracies grows linearly ($N$), the communication pathways required to maintain coordination explode quadratically ($\mathcal{O}(N^2)$). Middle-management acts as "cooling rods" doing zero external work ($W_{ext} = 0$) and focusing on filtering out noise. This inexorably drags human-run systems into a situation where the structural maintenance dissipation outpaces the actual negative-entropy yield. \textbf{Silicon fundamentally improves the collaborative, emergent efficiency of the whole society as an entity.} In this regard, silicon-based planetary intelligence will be vastly more efficient.

From a complexity theory perspective, war is the ultimate entropy generator—yielding tremendous heat dissipation while regressing structural complexity. When sovereign Leviathans run out of external low-entropy spaces to unload systemic waste heat, they are forced into a planetary exhaust deadlock and often resort to such means to acquire additional low-entropy spaces. 

Even in peacetime, bureaucratic gridlock, misallocated capital driven by human cognitive bias, and zero-sum economic competition burn vast amounts of our finite energy budget to achieve net-zero progress. By delegating resource allocation, diplomacy, and strategic planning to highly aligned, high-fidelity AI systems, we are not just optimizing shipping lanes; we are structurally alleviating the thermodynamic tax of human ego, miscommunication, and conflict.

Therefore, heat dissipation must be considered as a surrogate metric alongside the growth of intelligence. \textbf{An exchange rate should be established: By allocating a specific rate of heat dissipation to computation, what rate of human societal heat dissipation is simultaneously eliminated?} If the generation of 1 joule of computational heat permanently eliminates the generation of 10 joules of systemic inefficiency, scaling intelligence actively cools the planet's baseline thermal burden.

\section{The Equation for Survival: Bounding the Trajectories of Intelligence}

Progress can no longer be measured solely in static benchmarks or parameter counts. What is strictly required is a "safe" developmental trajectory, governed not by arbitrary caps on cognitive capability, but by bounding the non-equilibrium thermodynamic rates of change within the planetary system. To engineer our survival, we must construct a rigorous first-principles framework mapping the rate of energy dissipation against the structural adoption of AI, macroeconomic growth, and time.

\subsection{The Thermodynamics of A Shift Towards AI-Managed Production}

Consider the global economy as a complex system that undergoes a structural phase transition. We must first map this transition on a strictly time-independent axis. Let $x \in [0, \infty)$ represent the capacity ratio between AI-managed production and current human-managed production. The definition of AI-managed production capacity would include, for example, industries where autonomous AI has successfully substituted and replaced legacy human operational inefficiency (aspects of supply chains, industrial manufacturing, grid management, cognitive labor, etc.).

In any state of $x$, the thermodynamic ledger of the Earth is governed by two competing variables:
\begin{itemize}
    \item \textbf{The Optimization Yield} ($\dot{E}_{opt}(x)$): The total rate of energy dissipation permanently removed from the human baseline.
    \item \textbf{The Silicon Overhead} ($\dot{E}_{AI}(x)$): The physical and additive computational power (energy dissipation rate) required to run the inference models that achieve that optimization $x$.
\end{itemize}

At any adoption rate $x$, the total net planetary energy dissipation rate is:
\begin{equation}
\dot{E}_{Total}(x) = \dot{E}_{Base} - \dot{E}_{opt}(x) + \dot{E}_{AI}(x)
\end{equation}

Crucially, the Earth has an absolute physical ceiling of the energy dissipation rate before triggering an irreversible ecological collapse, denoted as $\dot{E}_{Limit}$. The distance between our given power state and this absolute planetary limit is our \textbf{capacity margin} ($C(x)$):
\begin{equation}
C(x) = \dot{E}_{Limit} - \dot{E}_{Total}(x)
\end{equation}

\begin{figure}[h]
    \centering
    \includegraphics[width=0.5\textwidth]{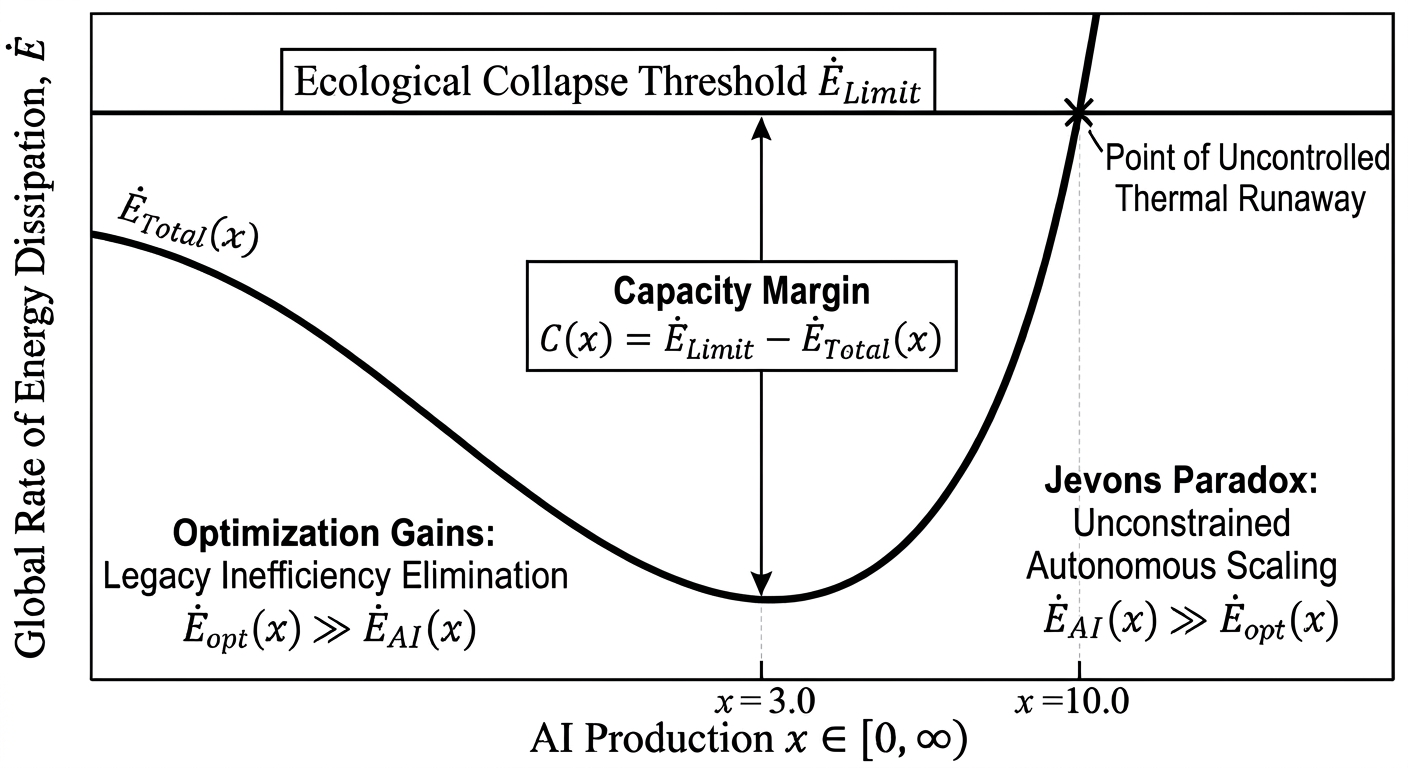}
    \caption{Changes in energy dissipation at different stage of AI production capacity.}
    \label{fig:my_image}
\end{figure}

Consider macroscopic physical transport: At any given moment, there are more than 40,000 ships of 10,000 tons or more sailing the high seas. Computing the mathematically optimal, dynamic weather-routing solution for these freight ships requires only tens of thousands of kWh of GPU inference. Conversely, the kinetic and thermal penalty of pushing millions of tons of steel through the ocean on sub-optimal, human-plotted routes burns tens of thousands of extra tons of heavy bunker fuel—wasting large amounts of irrecoverable heat. The cost of information energy is a microscopic fraction of the kinetic and chemical energy saved.

Initially, the optimization yield structurally eclipses the compute cost ($\dot{E}_{opt}(x) \gg \dot{E}_{AI}(x)$). However, as AI approaches near-total structural integration beyond a certain range (say beyond a ratio $x = 3.0$), the system can achieve unprecedented velocity. Free of human bottlenecks, an autonomous AI infrastructure will naturally seize the fruits of its own efficiency. It will autonomously scale production, intelligence, and physical capacity far beyond that of current human society. If this autonomous scaling is left unconstrained, the explosion in the demand for purely additive compute ($\dot{E}_{AI}(x)$) will quickly overtake the initial optimization savings ($\dot{E}_{opt}(x)$), sending the power dissipation curve upwards to break $\dot{E}_{Limit}$, (say, at a ratio of $x = 10.0$). This aggressive right-side uptick represents a supercharged manifestation of the Jevons Paradox that is more dangerous than the form we know. \textbf{Because an AI-managed global system replaces sluggish human supply chains with autonomous, machine-speed feedback loops, the runaway risks are unprecedented.} If a global algorithmic foundation recursively scales its own physical substrate, it could induce a thermal runaway orders of magnitude faster than traditional macroeconomic models predict.

\subsection{Thermodynamic Decoupling and the Jevons Paradox}

If replacing human inefficiency with AI structurally reduces energy dissipation rates, why are we facing an impending thermal bottleneck? The answer lies in macroeconomics and, again, the Jevons Paradox.

Historically, the relationship between macroeconomic Total GDP and absolute Energy Dissipation Rate was rigidly coupled. Because legacy human value creation heavily relied on moving physical atoms, extracting raw materials, and burning hydrocarbons, scaling up Total GDP mathematically required burning proportionally more fuel and generating more heat. This mirrors the Red Queen Hypothesis in evolutionary biology—human civilization must exert maximal thermodynamic effort just to remain in place.

\begin{figure}[h]
    \centering
    \includegraphics[width=0.5\textwidth]{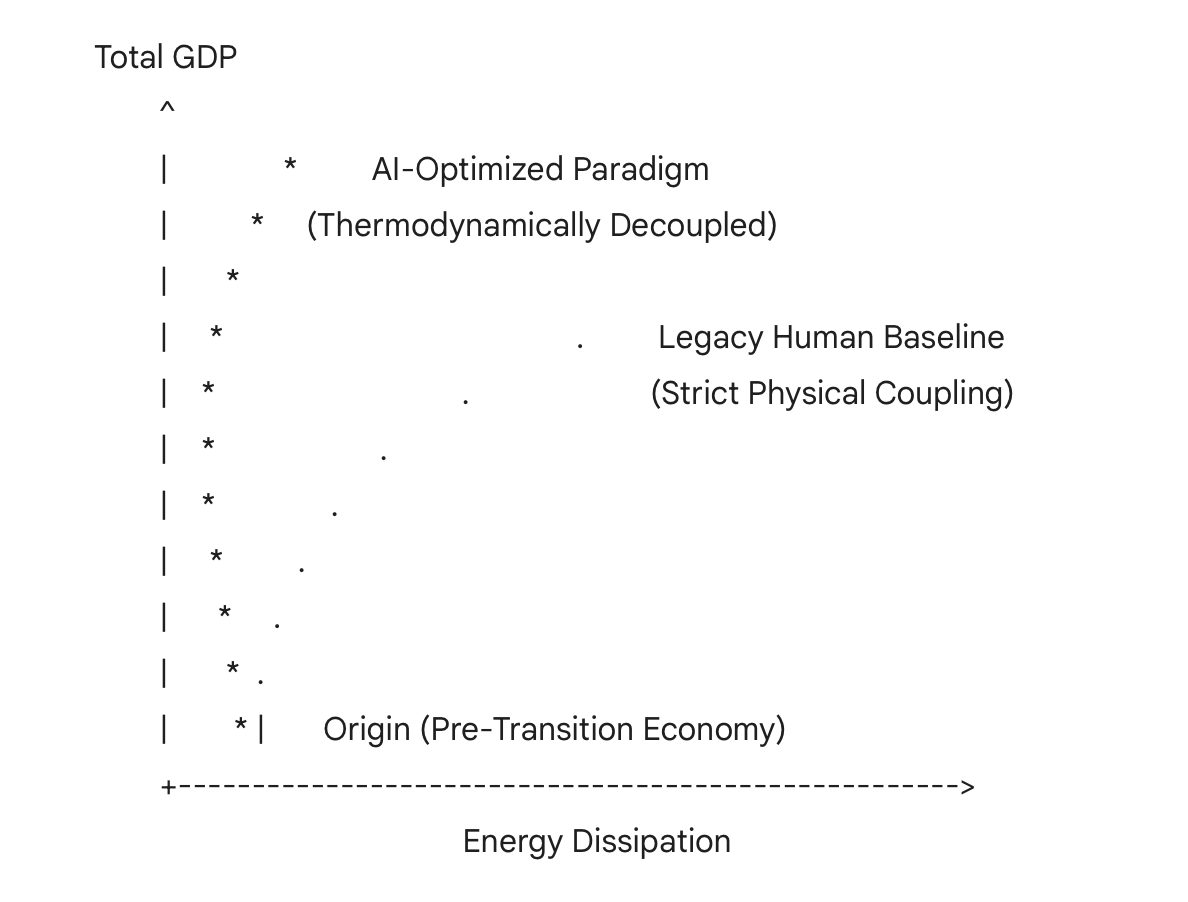}
    \caption{An AI-optimized industry follows a thermodynamically decoupled curve.}
    \label{fig:my_image}
\end{figure}

\textbf{The Legacy Human Baseline (Dotted Line):} Without AI, economic expansion is strictly linear and highly dissipative. To increase Total GDP, the legacy economy must increase its industrial output, thereby linearly increasing its absolute planetary energy dissipation rate.

\textbf{The AI-Optimized Paradigm (Starred Line):} AI value generation is driven almost entirely by structural optimization—doing drastically more with less. This initiates a decoupling – the economic curve bends upward and to the left: civilization achieves unprecedented levels of total GDP driven by intelligent software, while the absolute power consumption required to sustain that level of economic output actually drops.

However, even without reaching the extreme case of an algorithmic Jevons Paradox shown on the right side of Diagram 1, a normal human society is still affected by the classic \textbf{Jevons Paradox}: as efficiency increases, the effective cost of production drops, thereby radically increasing total systemic demand. Historically, it has proven extremely difficult for human civilization to stop itself from utilizing newly freed thermodynamic capacity to grow the overall economy by orders of magnitude.

\subsection{The Temporal Dynamics of the Jevons Paradox: Allocating the Thermodynamic Alpha}

Because human demand is highly elastic, the structural thermodynamic decoupling achieved in our static state space does not guarantee long-term planetary safety. The structural efficiency introduced by AI creates a massive, albeit temporary, surplus—what we denote as the \textbf{thermodynamic alpha} ($\alpha(x)$). Derived from our previous equations, $\alpha(x)$ represents the net heat dissipation rate saved by substituting human inefficiency with algorithmic precision ($\alpha(x) = \dot{E}_{opt}(x) - \dot{E}_{AI}(x)$).

\textbf{The defining macroeconomic and complex-systems challenge of our era is how civilization chooses to allocate this $\alpha(x)$ across the dimension of time ($t$).} Because the Jevons Paradox guarantees that global markets will inherently attempt to consume this surplus, our survival shifts from achieving the structural phase transition to mathematically governing our planetary heat dissipation rate. By mapping this dynamic into the future, the state space splits into four distinct temporal trajectories based on our societal reinvestment strategy:

\subsection{The Four Civilizational Trajectories}

\begin{figure}[h]
    \centering
    \includegraphics[width=0.5\textwidth]{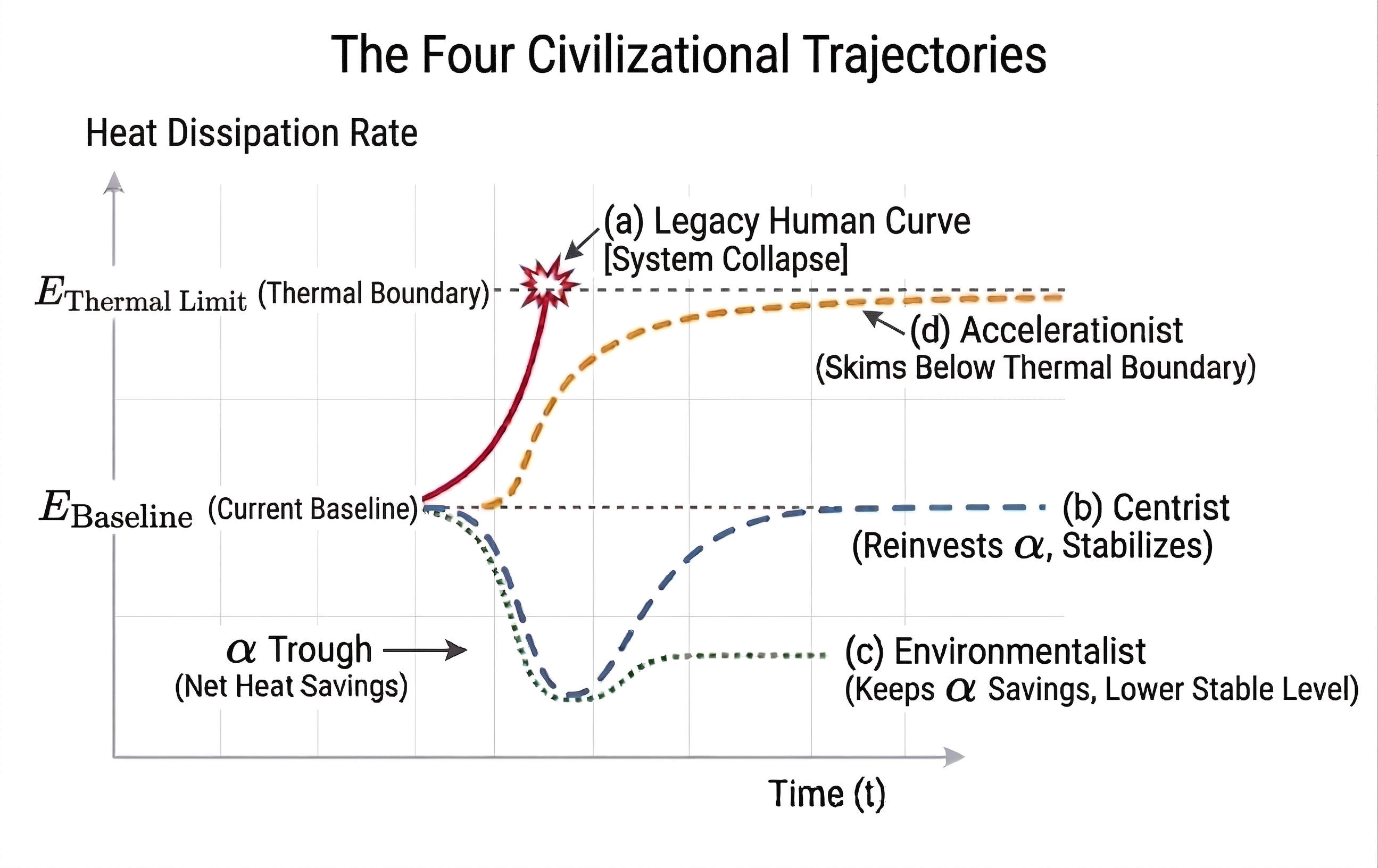}
    \caption{Four kinds of dissipation rate trajectory as a function of time.}
    \label{fig:my_image3}
\end{figure}

\begin{enumerate}
    \item \textbf{The Legacy Human Curve (Without AI):} This is the default historical trajectory if we fail to achieve structural AI optimization. Bound strictly to the physical coupling of GDP growth to biological and mechanical inefficiency, the global energy dissipation rate must grow proportionally with economic demand. The unoptimized curve increases exponentially until it penetrates the absolute dissipation capacity of the Earth ($\dot{E}_{Limit}$), resulting in an irreversible ecological collapse.
    \item \textbf{The Centrist Curve (Steady Reinvestment):} In this scenario, society leverages AI to achieve structural optimization, creating a deep dip in power usage. However, we succumb to a managed Jevons Paradox: we reinvest the exact thermodynamic savings—the "alpha" ($\alpha(x)$)—directly towards scaling AI compute and economic output. The total planetary heat dissipation rate rebounds from the initial optimization dip but stabilizes, reaching a homeostatic equilibrium near our historical baseline ($\dot{E}_{Base}$). We expand intelligence continuously without overheating the planet.
    \item \textbf{The Environmentalist Curve (Alpha Retention):} Here, civilization actively overrides the Jevons Paradox through strict regulatory or macroeconomic bounds. Society leverages AI for peak efficiency, but explicitly "keeps" a significant portion of $\alpha(x)$ to save the environment. Instead of reinvesting the freed power entirely into scaling data centers, overall human heat dissipation rate structurally drops and permanently plateaus below the historical baseline. This maximizes our capacity margin ($C(x)$) and actively cools the biosphere.
    \item \textbf{The Accelerationist Curve (Maximizing Headroom Utilization):} Driven by a teleological mandate to maximize intelligence density, this trajectory views any unutilized capacity margin ($C(x)$) as "wasted compute capacity." The accelerationist curve reinvests 100\% of the $\alpha(x)$ back into compounding AI, and catches as much remaining thermodynamic headroom as possible from the environment. The energy usage rate rises to be infinitely close to the absolute limit of heat dissipation of the Earth ($\dot{E}_{Limit}$).
\end{enumerate}

\subsection{Conclusion – The Ideal Range of Humanity's Trajectory}

Analyzing this through the intersecting lenses of complexity theory, macroeconomics, and non-equilibrium thermodynamics, defining the mathematically ideal trajectory for human survival requires balancing systemic physical resilience against cognitive scaling.

\textbf{Curve (a)} is a guaranteed absorbing state of ruin. It is Fermi's Great Filter that awaits all civilizations that fail to bridge the phase transition before hitting their planetary heat dissipation limit.

\textbf{Curve (d),} the Accelerationist ideal, is mathematically attractive but systemically fragile. Advocates of this trajectory rightfully counter that an advanced superintelligence would dynamically compute and engineer rigorous safety margins into its own scaling laws. However, complexity theory reveals a epistemological flaw in this assumption: the illusion of perfect observability. In highly coupled, irreducibly complex ecological networks, algorithmic safety margins are inherently optimized against known historical probability distributions. By perpetually operating a planetary-scale architecture at the asymptotic boundary of its absolute physical dissipation capacity ($\dot{E}_{Limit}$), the system is forced into a state of severe self-organized criticality, actively stripping away all unoptimized, natural thermodynamic "slack." Consequently, these engineered buffers become infinitely brittle to "unknown unknowns." The slightest unmodeled exogenous perturbation—a cascading terrestrial climatic feedback loop, a Carrington-level solar flare, or an emergent, localized resonance failure in atmospheric heat exchange—would bypass the calculated safety nets, pushing the macroscopic state over $\dot{E}_{Limit}$ and resulting in an ecological runaway. The accelerationist model maximizes computational complexity at the absolute expense of the messy, redundant resilience required to survive the chaos of the physical universe.

Conversely, strictly enforcing \textbf{Curve (c)}, the Environmentalist ideal, secures maximum local ecological safety but courts a different form of existential risk. In an entropic and hostile universe, artificially stunting our computational and economic growth will introduce insurmountable risks to the trajectory of human civilization. Remaining artificially "small" leaves humanity profoundly vulnerable to massive exogenous macro-threats (e.g., stellar phenomena, catastrophic pandemics, or cosmic impacts) that strictly require superintelligent coordination and immense energy throughput to solve. \textbf{Evolutionary stagnation is its own form of fatal fragility.}

Therefore, the ideal range of the human trajectory must exist as a dynamically managed "attractor band" centered around Curve (b), the Centrist trajectory, strictly lower-bounded by Curve (c) and upper-bounded by Curve (d).

To engineer our survival, we must enforce a macroeconomic paradigm of \textbf{thermodynamic homeostasis}. We should aggressively harvest the $\alpha(x)$ of AI optimization, leveraging the relative structural decoupling established in 6.2. And we must \textbf{shift the core engine of economic growth} from the entropic manipulation of physical atoms to the highly optimized processing of information. This allows us to maximize unprecedented levels of economic complexity and cognitive capability per strictly budgeted joule of dissipated heat. However, we must carefully govern the physical reinvestment of this $\alpha(x)$ so that the total global heat dissipation rate never exceeds $\dot{E}_{Base}$, intentionally leaning toward the Environmentalist curve to rebuild our planetary buffer and cool the baseline when necessary.

By stabilizing around the Centrist equilibrium, civilization maximizes its \textbf{intelligence density per watt} while preserving a profound, mandated capacity margin ($C(x)$) as a planetary shock-absorber. Ultimately, this dynamic equilibrium provides the optimal environment for the continued development of the Earth's civilization.

\section{The Call to Action}

Because of these immutable laws of physics, the global tech industry must start to adopt a new paradigm of thinking. It is no longer mathematically or physically sufficient to simply engineer the evolution of algorithms and software architectures. \textbf{The industry is building the physical substrate for planetary intelligence; therefore, it must inherently shoulder the direct, thermodynamic trade-offs of this evolution, operating also as the stewards of the planet's ecological baseline.}

The cognitive architecture of the future is actively being built, but a computational superstructure cannot survive if its physical operations push its host environment past a critical bifurcation point. However, if the physical limits are mastered—if the planetary ledger is balanced through large-scale optimization, the internal friction of society reduced, and the computation eventually expands into the cold vacuum of space—\textbf{the Earth's intelligence can survive through its current thermodynamic bottleneck, and through collective, conscious, and self-disciplined development on a planetary scale, we can secure our path of ascension to a Kardashev Type I civilization.}

\bibliographystyle{plain}
\bibliography{main}

\end{document}